\documentclass[pdflatex,sn-mathphys-num]{sn-jnl}
\usepackage{graphicx}
\usepackage{multirow}
\usepackage{amsmath,amssymb,amsfonts}
\usepackage{amsthm}
\usepackage{mathrsfs}
\usepackage[title]{appendix}
\usepackage{xcolor}
\usepackage{textcomp}
\usepackage{manyfoot}
\usepackage{booktabs}
\usepackage{algorithm}
\usepackage{algorithmicx}
\usepackage{algpseudocode}
\usepackage{listings}
\usepackage{physics}

\theoremstyle{thmstyleone}

\theoremstyle{thmstyletwo}

\theoremstyle{thmstylethree}

\begin{document}

\title[Article Title]{First-Principles Calculation of Spin-Relaxation Due to Alloy and Electron-Phonon Scattering in Strained GeSn}

\author*[1]{\fnm{Kevin} \sur{Sewell}}\email{kevin.sewell@tyndall.ie}

\author*[1]{\fnm{Felipe} \sur{Murphy-Armando}}\email{philip.murphy@tyndall.ie}

\affil*[1]{\orgdiv{Photonics Theory Group}, \orgname{Tyndall National Institute, University College Cork}, \orgaddress{\street{Lee Maltings Complex Dyke Parade}, \city{Cork}, \postcode{T12 R5CP}, \state{Cork}, \country{Ireland}}}

\abstract{GeSn has emerged as a promising material for spintronics due to its long spin-lifetime, compatibility with silicon technology, high mobility and tunable electronic properties. Of particular interest is the transition from an indirect to a direct band gap with increasing Sn content, which enhances optical properties, electron transport and we find also affects spin transport behaviour, which is critical for spintronics applications. We use first-principles electronic-structure theory to determine the spin-flip electron-alloy scattering parameters in n-type GeSn alloys. We also calculate the previously undetermined intervalley electron spin-phonon scattering parameters between the $L$ and $\Gamma$ valleys. These parameters are used to determine the electron-alloy and electron-phonon scattering contributions to the n-type spin-relaxation of GeSn, as a function of alloy content and temperature. As in the case of phonon scattering, alloy scattering reduces the spin-relaxation time. However, switching the spin transport from the typical $L$ valley of Ge to the $\Gamma$ valley by sufficient addition of Sn, the relaxation time can be substantially increased. For unstrained, room temperature GeSn, we find a Sn concentration of at least $10\%$ is required to achieve a spin-relaxation time greater than Ge, with $17\%$ Sn needed to increase the spin-relaxation time from the nanosecond range to the microsecond range. At low temperatures (30K), adding $10\%$ Sn can increase the spin-relaxation time from $10^{-7}$s to 0.1s. Applying biaxial tensile strain to GeSn further increases the spin-relaxation time and at a lower Sn content than in unstrained GeSn.}


\maketitle

\section{Introduction}

Spintronics has the potential to revolutionise electronics by enabling the development of faster, smaller, and more energy-efficient devices compared to conventional electronics, holding promise for applications in (quantum) computing, storage, and communication technologies \cite{awschalom2007challenges}. The ultimate aim of spintronics is to manipulate the spins of electronic charge as carriers of information rather than charge current. Spin-relaxation, a key feature of spin polarisation, describes the decay rate of a non-equilibrium spin-polarised population. This process is crucial for spintronics, as reliable spin-based information transfer requires protection against depolarisation from spin scattering. Consequently, materials with long spin lifetimes are sought after for applications such as spin-based transistors, quantum computing and spin interconnects. In contrast, other spintronics applications such as ultra-fast optical switches require short spin lifetimes \cite{marchionni2021inverse,vzutic2004spintronics}.

In centrosymmetric crystals such as silicon (Si) and germanium (Ge), spin-relaxation occurs mainly through the Elliott-Yafet (EY) mechanism \cite{elliott1954theory,yafet1963g}, where spin decoherence can be mediated by electron-phonon (e-ph) and/or electron-defect interactions. Si is an attractive spintronics material due to its relatively long spin-phonon-relaxation time, which lies in the nanosecond (ns) range at room temperature \cite{elliott1954theory,appelbaum2007electronic,cheng2010theory}. In comparison, Ge has a higher carrier mobility shorter spin-relaxation time, which also lies in the ns range \cite{li2012intrinsic,tang2012electron} (see Hamaya et al \cite{hamaya2018spin} for a detailed list of previously determined spin-relaxation times of electrons and holes in Ge at various temperatures and doping concentrations). Ge is a promising candidate for spin-to-charge conversion \cite{dushenko2014spin,dushenko2015experimental,oyarzun2016evidence,bottegoni2017spin}, and likely candidate material in spintronics \cite{hoffmann2015opportunities,de2019spin}. Typical spin-diffusion lengths in Ge are on the order of a few micrometers \cite{marchionni2021inverse,zucchetti2019doping,hamaya2018spin}. In comparison, direct-gap III-V materials InAs, InSb and GaSb, which are governed by the D'yakonov-Perel mechanism \cite{dyakonov1971current}, have spin-relaxation times in the picosecond range \cite{song2002spin,hautmann2009temperature,boggess2000room,bandyopadhyay2014coherent,murzyn2003suppression}, while the spin-relaxation time for GaAs lies between the picesecond and nanosecond range \cite{seymour1980time,ohno1999spin,oertel2008high}. In 2019, Cesari et al. \cite{de2019spin} measured the carrier lifetime to be in the nanoseconds range in strained $Ge_{1-x}Sn_x$ on Si semiconductors, hypothesising that $\alpha$-Sn could further enrich the spin-dependent phenomena observed in Ge, making it a potential candidate material in spin interconnects \cite{dery2011silicon,vzutic2019proximitized}. In addition, GeSn is already viewed as an attractive alternative to Si or Ge in electronics, having been demonstrated in a wide range of optoelectronic applications  \cite{cardoux2022room,huang2019electrically,chang2017room,kim2022improved,ojo2022silicon,marzban2022strain,wirths2015lasing,atalla2022high,talamas2021cmos,tran2019si,zhang2018gesn,miller2009device,liu2018chip}, as well as potentially possessing an electron mobility over $25$ times higher than that of Ge \cite{sewell2025first}.

Despite its obvious potential, research on GeSn alloys, particularly in spintronics, remains in its early stages for a variety of reasons. The solubility of Sn in Ge is low \cite{ayan2019strain,thurmond1953equilibrium} due to the large lattice mismatch between the two materials. Consequently, Sn segregates out of the Ge crystal without out-of-equilibrium processing \cite{li2013strain,cai2022thickness}. At present, Sn concentrations as high as $17-25\%$ Sn has been successfully incorporated into Ge \cite{diao2025magnetron,assali2019enhanced,schwarz2020alloy} through various growth techniques. While Ge is a indirect-gap material, with its conduction band minima at the $L$ valley, unstrained GeSn undergoes an indirect-to-direct transition at Sn compositions 7-9\% Sn concentration\cite{sewell2025first,jiang,ghetmiri2014direct,gallagher2014compositional}, or around $4.7\%$ Sn with $1\%$ biaxial tensile strain \cite{sewell2025first}. As a result, the total spin-relaxation time $\tau_s$ for GeSn depends not just on intravalley and intervalley scattering between the $L$ valleys, but also interactions between the $L$ valleys and the $\Gamma$ valley, as well as intravalley scattering at the $\Gamma$ valley. In addition, the effect of alloy-impurities on spin-flips is not well known, although it is known that in electron transport alloy scattering is more dominant than e-ph scattering in GeSn at sufficient Sn concentrations, even at room temperature \cite{sewell2025first}.

In this work we determine the spin-relaxation time in GeSn as a function of Sn concentration, temperature and strain. We use first-principles methods to calculate the spin-alloy scattering parameters for GeSn. These parameters are then used to determine the spin-alloy contributions to the spin-relaxation time. We also implement first-principles methods to calculate the electron-spin-phonon contribution to the spin-relaxation time. Finally, we use these findings (along with our findings in Ref. \cite{sewell2025first}) to calculate the spin-diffusion length as a function of Sn concentration at fixed temperatures. By comparing our findings to those for Ge \cite{li2012intrinsic,tang2012electron}, we can understand whether or not spin-alloy scattering offsets the advantage of introducing Sn to Ge for the purpose of its high spin orbit coupling (SOC).


\section{Method}


To calculate the thermal averaged spin-relaxation time $\langle\tau^\alpha\rangle$ at Brillouin zone (BZ) valley $\alpha$, we use $\langle\tau^\alpha\rangle=\frac{1}{\langle R^\alpha\rangle}$, where $\langle R^\alpha\rangle$ is the thermal average of the scattering rate $R^\alpha(E)$:
\begin{equation} \label{eq:thav}
    \langle R^\alpha \rangle = \frac{\int R^\alpha(E)  \ \frac{\partial f}{\partial E} \ \rho^\alpha(E) \ dE}{\int \frac{\partial f}{\partial E} \ \rho^\alpha(E) \ dE}
\end{equation}
    
where $f(E)$ is the Fermi-Dirac distribution function, which simplifies to Boltzmann distribution at room
temperature, and $\rho^\alpha(E)$ is the non-parabolic density of states near the band minima $\alpha$, which is calculated using methods described in Ref. \cite{sewell2025first}. The conduction band minima energy $E_c^\alpha$ (or bandgap) are calculated using the GW approximation \cite{GW} and corrected empirically for temperature using Varshni's empirical expression \cite{varshni1957comparative}. The alloy content is represented by the Virtual Crystal Approximation (VCA), where the bandgap at each atom is taken to be $E_x^{VCA}=(1-x)E^{Ge}+xE^{Sn}+b^{GeSn}x(1-x)$. The bowing parameter $b$ is taken from Mukhopadhyay et al \cite{mukh}. Our calculated bandgaps and effective masses can be found in Ref. \cite{sewell2025first}. We only consider carriers populating the $L$- and $\Gamma$-valleys, since under our conditions the contribution from other valleys (i.e. $\Delta$ conduction valleys) is negligible in GeSn. Consequently, only the intravalley and intervalley scattering parameters between $L$ and $\Gamma$ affect our calculation of the spin-relaxation time $\langle\tau\rangle$, whereby:
\begin{equation} \label{eq:tausum}
\langle\tau\rangle=r_L\langle\tau_L\rangle+r_\Gamma\langle\tau_\Gamma\rangle
\end{equation}

where both $\langle\tau_L\rangle$ and $\langle\tau_\Gamma\rangle$ include spin-relaxation as a result of both e-ph and alloy scattering. In Eq. \ref{eq:tausum}, the parameters $r_L$ and $r_\Gamma$ are the carrier concentration ratios of the $L$ and $\Gamma$ bands, respectively.

\subsection{Spin-alloy scattering}

The spin-momentum scattering rate $R^\alpha(E)$ from Eq. \ref{eq:thav} is determined by a combination of alloy scattering and e-ph scattering. For an electron with initial and final spin polarisations $s_1$ and $s_2$ respectively, its alloy scattering rate from BZ valley $\alpha$ to $\beta$ due to alloy disorder in the random substitutional alloy \cite{fma06} is given by:
\begin{equation}
R_{\alpha\beta}^{s_1s_2}(E)=\frac{2\pi}{\hbar}x(1-x)\frac{a_0^3}{8} \sum_{\beta} |\langle V_{\alpha\beta,s_1s_2} \rangle|^2  \rho^{\beta}(E)
\end{equation}

where $x$ is the Sn concentration, $a_0$ is the lattice constant and $\langle V_{\alpha\beta,s_1s_2} \rangle$ is the scattering matrix:
\begin{align} 
\left\langle V_{\alpha\beta,s_1s_2} \right\rangle 
&= \left\langle V_{\alpha\beta,s_1s_2}^{Sn} \right\rangle - \left\langle V_{\alpha\beta,s_1s_2}^{Ge} \right\rangle \nonumber \\
&=N( \bra {\psi_{\alpha,s_1}} {\Delta V^{Sn}} \ket{\phi_{\beta,s_2}} - \bra{ \psi_{\alpha,s_1} }{ \Delta V^{Ge} } \ket{ \phi_{\beta,s_2} }) \label{pot}
\end{align}

In Equation \ref{pot}, $N$ is the number of atoms, $\psi$ represents the Bloch state of the periodic host lattice and $\phi$ is the exact eigenstate of the perturbing potential $\Delta V^A$, which is caused by the substitution of atom $A$ into the periodic host. Each wavefunction $\ket{\psi}=(\psi_\Uparrow,\psi_\Downarrow)^T$ can be decomposed into effective up and down spin-states $\psi_\Uparrow$ and $\psi_\Downarrow$, which diagonalise the spin operator $S^\alpha=\frac{\hbar}{2}\sigma^\alpha$ in the Kramers degenerate subspace \cite{park2020spin}, where $\sigma^\alpha$ denotes the Pauli matrices corresponding to the Cartesian components $\alpha=x,y,z$:
\begin{align}
    \bra{\psi_\Uparrow} S_\alpha \ket{\psi_\Uparrow}&= -\bra{\psi_\Downarrow} S_\alpha \ket{\psi_\Downarrow} \\
    \bra{\psi_\Uparrow} S_\alpha \ket{\psi_\Downarrow}&= 0
\end{align}

The effective spin states are calculated from the spin matrix $S(k)$, which represents the spin operator in the wave function basis. The matrix elements of the spin operator in the Bloch state basis are given by $\sigma_{m,s_1 ; n,s_2}(k) = \bra{\psi_{m,k,s_1}} \sigma^\alpha \ket{\psi_{n,k,s_2}}$. Here, $\ket{\psi_{n,k,s_2}}$ denotes the Bloch state at wavevector $k$, band index $n$, and spin $s_2$. We take the spin quantisation axis to be aligned with the z-axis. We separately diagonalise each degenerate subspace in S at each k-point, obtaining the unitary matrices $D_k$ that make each of the subspaces in $D_kS(k)D^\dag_k$ diagonal, with eigenvalues equal to the effective spin \cite{park2020spin}. The spin-flip alloy matrix elements $V_{\alpha\beta,\uparrow\downarrow}$ are then computed using Eq. \ref{pot} for all pairs of states with opposite effective spin. The boundary condition $\psi(\Vec{r})= \phi(\Vec{r})$ is imposed when $\Vec{r}$ is far from the type-A atom. In an infinite crystal, $\Delta V^A$ tends to $0$ at a large distance from the impurity site. In a supercell calculation, the zero of the potential energy is arbitrary, so a physically well-defined method must be developed to compare the potentials of the supercell with N host atoms and the supercell with one type-A atom and $N-1$ host atoms. We address this using methods described in detail in Refs. \cite{sewell2025first,fma08}. A $64$-atom supercell was used to find the scattering matrices between degenerate Bloch states, which was large enough to allow for realistic structural relaxation of the host around the substitutional Ge or Sn atom. We implement lattice parameters calculated in Ref. \cite{sewell2025first}. The first-principles calculations are performed using Hartwigsen-Goedecker-Hutter \cite{hartwigsen1998relativistic} plane-wave pseudopotentials in all of our Density Functional Theory calculations with ABINIT code \cite{ab}. The energy cut-off was $950eV$ and 6x6x6 Monkhorst-pack k-meshes were generated for the first BZ integration. We apply the local density approximation for the exchange-correlation functional.

\subsection{Electron spin-phonon scattering}

Previous studies \cite{li2012intrinsic,tang2012electron} show that spin-relaxation in Ge is primarily governed by intra- and intervalley e-ph spin-scattering between the $L$ valleys. We calculate the spin-relaxation time due to electron spin-phonon scattering in these valleys using phonon parameters for Ge calculated by Tang et al. \cite{tang2012electron}, assuming weak dependence on Sn concentration at low Sn compositions.

However, in GeSn, both (a) intravalley scattering at the $\Gamma$ valley and (b)intervalley scattering between $L$ and $\Gamma$ also dictate the behaviour of GeSn's spin-relaxation time and the parameters are not yet available in the literature. 
(a) We follow Tamborenea et al \cite{tamborenea2003spin} to calculate the spin-relaxation time for intravalley electron spin-phonon scattering at the $\Gamma$ valley. They describe the spin-flip transition rate using Fermi's Golden Rule and the EY mechanism, expressing the matrix element as a product of the spin-independent e-ph matrix element $V$ and the overlap integral $\left<\mathbf{k'}\downarrow|\mathbf{k}\uparrow\right>$, such that the scattering rate $R_{\mathbf{k}\uparrow,\mathbf{k'}\downarrow}$ is given by:
\begin{equation}
    R_{\mathbf{k}\uparrow,\mathbf{k'}\downarrow}=\frac{2\pi}{\hbar}\delta\left(E_{\mathbf{k'}}-E_{\mathbf{k}}\right) V\left(\mathbf{k}-\mathbf{k'}\right)^2 \left|\left<\mathbf{k'}\downarrow|\mathbf{k}\uparrow\right>\right|^2
\end{equation}

We calculate the scalar product $\left<\mathbf{k'}\downarrow|\mathbf{k}\uparrow\right>$ by calculating the wavefunction overlaps of a dense set of k-points near the $\Gamma$ valley. Simple integration over $\mathbf{k}$ and $\mathbf{k'}$ obtains an expression of the spin-relaxation time due to intravalley scattering at the $\Gamma$ point.

(b) We use the frozen phonon supercell approach to calculate the deformation potentials for intervalley scattering between the $L$ and $\Gamma$ bands. We use long $4$-atom supercells to obtain the energy splittings and wavefunctions due to a phonon wave-vector $\mathbf{q}=\mathbf{k}_L-\mathbf{k}_{\Gamma}$ in the $[111]$-direction. The phonon frequencies ($\hbar\omega_\mathbf{q}$) were previously calculated \cite{murphy2011giant,tyuterev2011ab} using density functional perturbation theory (DFPT) with ABINIT, and correspond to the 6 different polarisations of an L-phonon. We set the phonon amplitude to $A=0.0025$ and $A=0.005$ to ensure convergence.



\section{Results}

\subsection{Spin-Alloy Scattering Parameters}

Calculations of the spin-alloy scattering matrix were performed using supercells with a VCA host at $0\%$, $6\%$, $12\%$ and $18\%$ Sn content. We observed that $V_{\alpha\beta}^{\uparrow\downarrow}$ largely exhibits a weak quadratic dependence on the Sn concentration, as we show in Table \ref{tab:ieflips}.

\begin{figure}[h]
    \centering
    \includegraphics[width=0.63\linewidth]{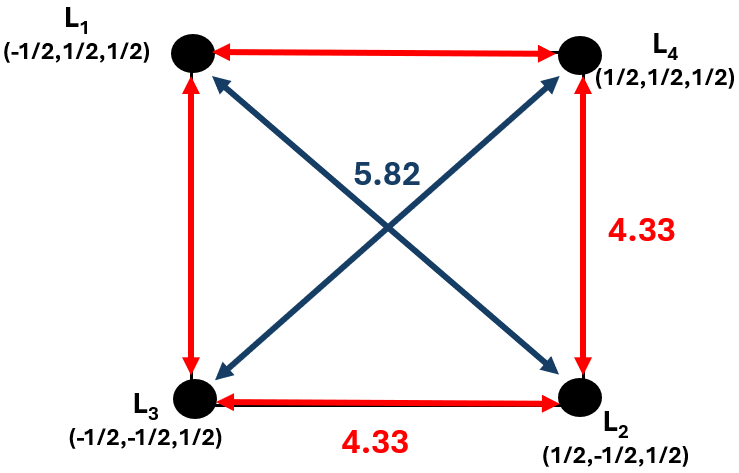}
    \caption{Intervalley spin-alloy scattering parameters between the L valleys of GeSn (in meV) with spin-polarisation in the z-direction.}
    \label{fig:Lflip}
\end{figure}

We find that spin-alloy scattering only occurs via intervalley scattering, and that spin-conserving scattering dominates momentum relaxation, as is the case in electron-phonon scattering\cite{li2012intrinsic}. Table \ref{tab:ieflips} shows the Sn-dependence of the intervalley spin-flip scattering matrix elements (in meV) for $Ge_{1-x}Sn_x$. Regarding intervalley scattering between the L valleys, the superscript $xy$ indicates that both the initial and final valleys are in the x-y plane, while the superscript $z$ indicates that the initial and final valleys are separated along the z direction. The spin-flip scattering parameter $V_{LL}^{z}$ is larger than $V_{LL}^{xy}$. This difference is highlighted in Fig. \ref{fig:Lflip}, where we show the scattering parameters calculated using a supercell with Ge as the host atom. $V_{LL}^{z}$ is denoted using blue arrows, while $V_{LL}^{xy}$ is denoted using red arrows.
We find that spin-alloy scattering is on the same order of magnitude as spin-phonon parameters at room temperature \cite{li2012intrinsic,tang2012electron}. Moreover, spin-flips between $L$ and $\Gamma$ are over an order of magnitude weaker than $L-L$ scattering, resulting in spin-flips due to alloy scattering being more prominent in high-Ge GeSn compared to direct-gap GeSn.

\begin{table}[h]
\centering
\renewcommand{\arraystretch}{2} 
\setlength{\tabcolsep}{7pt} 
\begin{tabular}{cc}
\hline
\textbf{Symbol} & \textbf{Value (meV)} \\ \hline
$V_{\Gamma L}$   & $0.13995 + 0.799167 x + 0.625 x^2$      \\ \hline
$V_{LL}^z$    & $5.8175 + 15.625 x - 7.63889 x^2$      \\ \hline
$V_{LL}^{xy}$  & $4.334 + 12.15 x - 4.16667 x^2$      \\ 
\hline
\end{tabular}
\caption{Calculated intervalley spin-alloy scattering matrix elements $V_{\alpha\beta}^{\uparrow\downarrow}$
(in meV) in $Ge_{1-x}Sn_x$, with spin-polarisation in the z-direction.}
\label{tab:ieflips}
\end{table}

We observe no spin-flip events resulting from intravalley alloy scattering processes. This indicates that intravalley scattering preserves the spin orientation of the electrons. Consequently, the intravalley scattering parameters that describe electron dynamics, specifically those related to momentum relaxation, are unaffected by the inclusion of SOC. In other words, since no spin-flip transitions occur within the same valley, the parameters governing intravalley scattering with spin conservation (found in Ref. \cite{sewell2025first}) are identical to those obtained when SOC effects are neglected.

\subsection{Electron Spin-Phonon Scattering Parameters}

Using the frozen phonon approach, we calculate all relevant phonon energies and deformation potentials for each of the six $L$-phonon modes are shown in Table \ref{tab:phie}. The corresponding matrix elements are calculated using these values. Our spin-conserving intervalley scattering parameters, also shown in Table \ref{tab:phie}, were found to be in excellent agreement with Ref. \cite{murphy2023electronic}.

\begin{table}[h]
\centering
\renewcommand{\arraystretch}{1.8} 
\setlength{\tabcolsep}{6pt} 
\begin{tabular}{ccccccc}
\hline
\textbf{Mode} & $1$ & $2$  & $3$  & $4$  & $5$  & $6$    \\ \hline
    $\mathbf{\hbar\omega_q}$ \textbf{(meV)} & $5.83$ & $5.83$  & $21.56$ & $23.23$ & $27.69$ & $27.69$ \\ \hline
$\mathbf{D_{L\Gamma}^{\uparrow\downarrow}}$ $\mathbf{(eV/\AA)}$ & $0.0102$   &  $0.0204$  & $9.805$$\times$$10^{-5}$ & $9.804$$\times$$10^{-5}$  & $0.02163$ & $0.02145$ \\ \hline
$\mathbf{D_{L\Gamma}^{\uparrow\uparrow}}$ $\mathbf{(eV/\AA)}$ & $0.013$   &  $0.009$  & $2.03$ & $2.03$  & $8$$\times$$10^{-4}$ & $0.012$ \\ \hline
\end{tabular}
\caption{Energies $\mathbf{\hbar\omega_q}$ (in meV), spin-flip deformation potentials $D_{L\Gamma}^{\uparrow\downarrow}$ and spin-conserving deformation potentials $D_{L\Gamma}^{\uparrow\uparrow}$ (in $eV/\AA$) for all phonon modes relevant for scattering between the $\Gamma$ and $L$ bands in $Ge_{1-x}Sn_x$.}
\label{tab:phie}
\end{table}

\subsection{Spin-Relaxation of GeSn}

\begin{figure}[h]
    \centering
    \includegraphics[width=\linewidth]{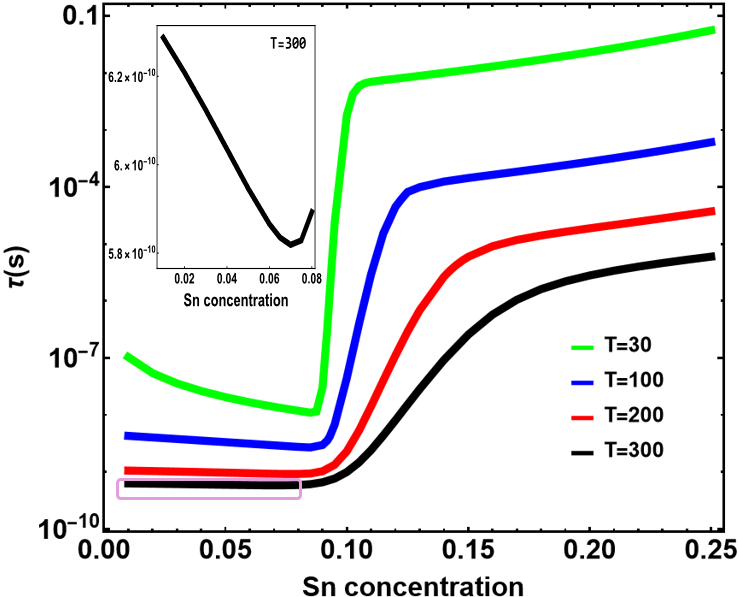}
    \caption{Intrinsic spin-relaxation time of GeSn at temperatures T=300K (black), T=200K (red), T=100K (blue) and T=30K (green), as a function of Sn concentration. We include an inset of the spin-relaxation time at 300K at low Sn concentration to highlight its decrease with alloy content.}
    \label{fig:tauSn}
\end{figure}

Figure \ref{fig:tauSn} shows the calculated intrinsic spin-relaxation time $\tau$ (given in seconds) at different fixed temperatures (T=$30$,$100$,$200$,$300$K) as a function of Sn concentration. At any fixed temperature, as Sn is introduced to Ge, the relaxation time initially decreases due to alloy scattering, especially at lower temperatures, where alloy scattering is more dominant than e-ph scattering. As the Sn concentration increases and the indirect-to-direct transition occurs, the spin-relaxation time increases by several orders of magnitude as electrons populate the $\Gamma$ valley. This can be attributed to the minimal spin-alloy scattering in direct-gap GeSn, along with reduced intravalley electron-spin phonon scattering at the $\Gamma$ valley compared to the $L$ valley, a result of the lower density of states in the $\Gamma$ band.

\bigskip

Alloying Ge with Sn results in an increase in the the spin-relaxation time from below the nanosecond range to the microsecond range (at a Sn concentration between $17\%$ and $25\%$). At lower temperatures, the electron distribution becomes steeper. This explains the sharp rise in the spin-relaxation time at the indirect-to-direct bandgap transition (at $8.4\%$ Sn concentration). The reduced electron spin-phonon scattering at lower temperatures leads to a higher relaxation time, approaching $0.1$ seconds at $25\%$ Sn concentration at $30$K.

\bigskip

\begin{figure}[h]
    \centering
    \includegraphics[width=0.8\linewidth]{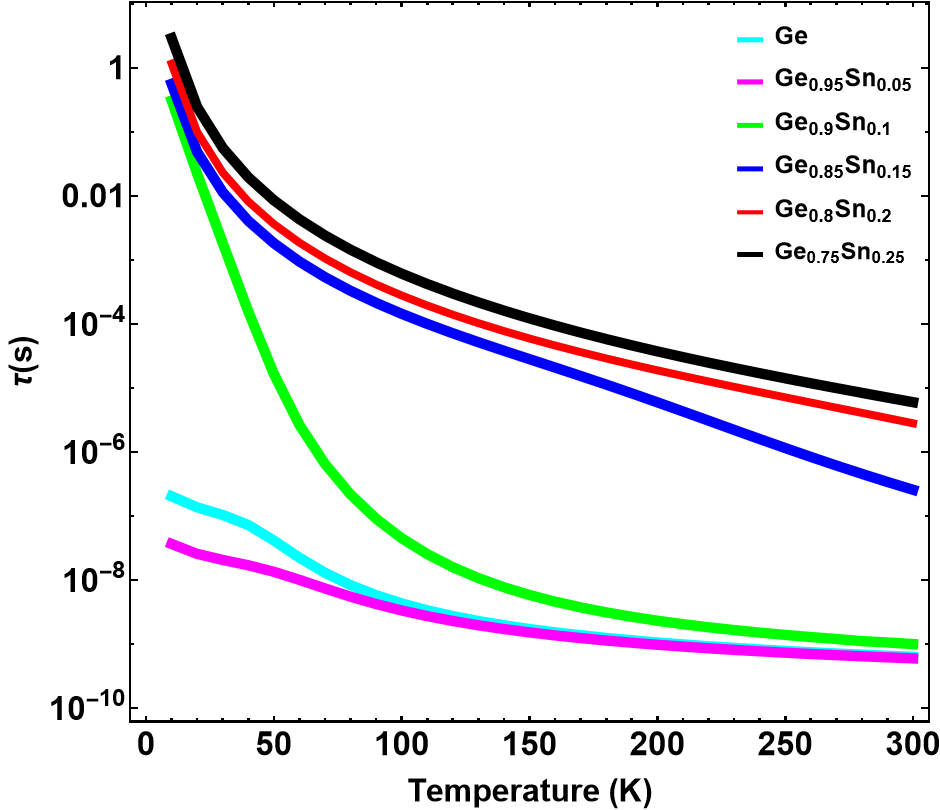}
    \caption{Intrinsic spin-relaxation time of Ge (light blue) and GeSn at Sn concentrations of $5\%$ (purple), $10\%$ (green), $15\%$ (dark blue), $20\%$ (red) and $25\%$ (black), as a function of temperature.}
    \label{fig:tauT}
\end{figure}

In Fig. \ref{fig:tauT}, we show the spin-relaxation time against temperature at Sn concentrations of up to $25\%$. In general, it is clear that spin-relaxation decreases as temperature increases due to increased electron spin-phonon scattering. Our values for Ge (light blue) are in good agreement with calculations by Tang et al \cite{tang2012electron}, where $\tau\approx0.5$ ns at room temperature, and approaches the microsecond range as $T\rightarrow0$. Above $100$K, at least $15\%$ Sn is required to increase the spin-relaxation time by several orders of magnitude. At low temperatures however, the spin-relaxation time can be increased by up to six orders of magnitude with the introduction of $10\%$ Sn. Due to the change in the behaviour of the distribution function with temperature, the spin-relaxation time for Ge$_{0.9}$Sn$_{0.1}$ varies from the nanosecond range to the second range between 30-300K. Beyond $15\%$ Sn concentration, there is little advantage in adding further Sn to increase the relaxation time, especially at lower temperatures.

\bigskip

\subsubsection{The Effect of Strain}

The behaviour of the spin-relaxation time of GeSn under biaxial tensile strain is analogous to that of unstrained GeSn. The
difference is that strain induces an indirect-to-direct bandgap transition at lower Sn concentrations; the consequence of which is a higher spin-relaxation time at any given temperature. Figure \ref{fig:taustrain} below shows the spin-relaxation time of unstrained (solid) and $1\%$ biaxially tensile strained (dashed) GeSn at $T=300K$ and $30K$ as functions of Sn concentration. The figure shows that at room temperature, beyond $5\%$ Sn concentration, $1\%$ biaxial tensile strain can increase the spin-relaxation time by almost two full orders of magnitude. At $30K$, adding $6.7\%$ Sn to $1\%$ strained GeSn is equivalent to introducing $10-15\%$ Sn to unstrained GeSn to increase the spin-relaxation time. Between $6-9\%$ Sn, $1\%$ strain increases the spin-relaxation time by almost six orders of magnitude at $30$K.

\bigskip

\begin{figure}[h!]
    \centering
    \includegraphics[width=0.8\linewidth]{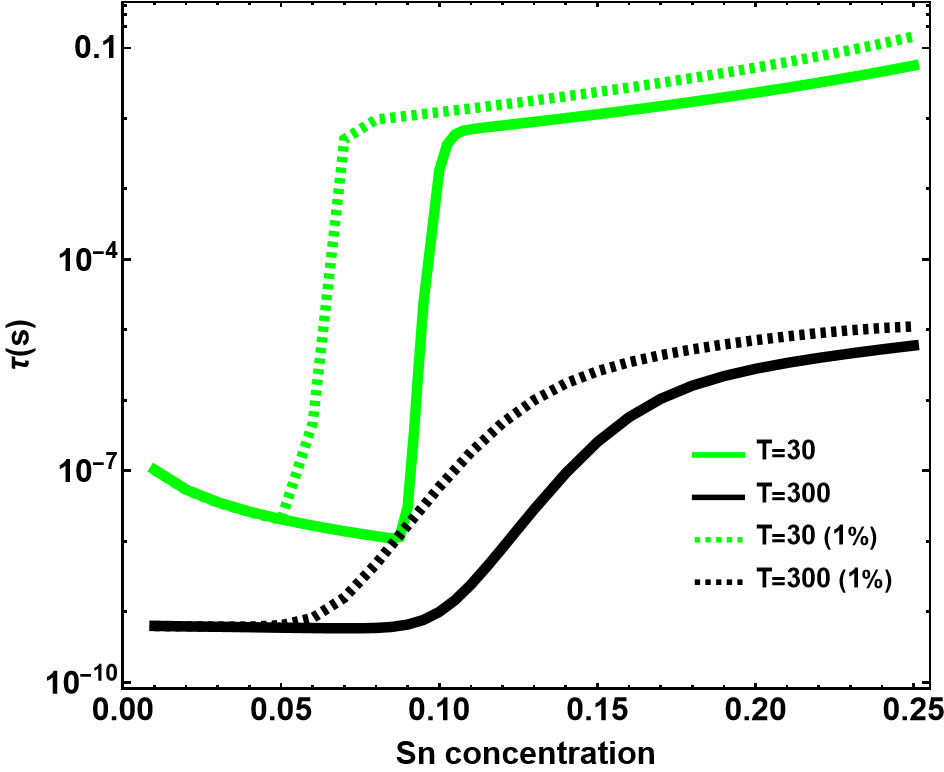}
    \caption{Spin-relaxation time of unstrained (solid) and $1\%$ biaxially tensile strained (dashed) GeSn at $300$K (black) and $30$K (green), as a function of Sn concentration.}
    \label{fig:taustrain}
\end{figure}

\subsubsection{Spin Diffusion Length}

\begin{figure}[h!]
    \centering
    \includegraphics[width=0.8\linewidth]{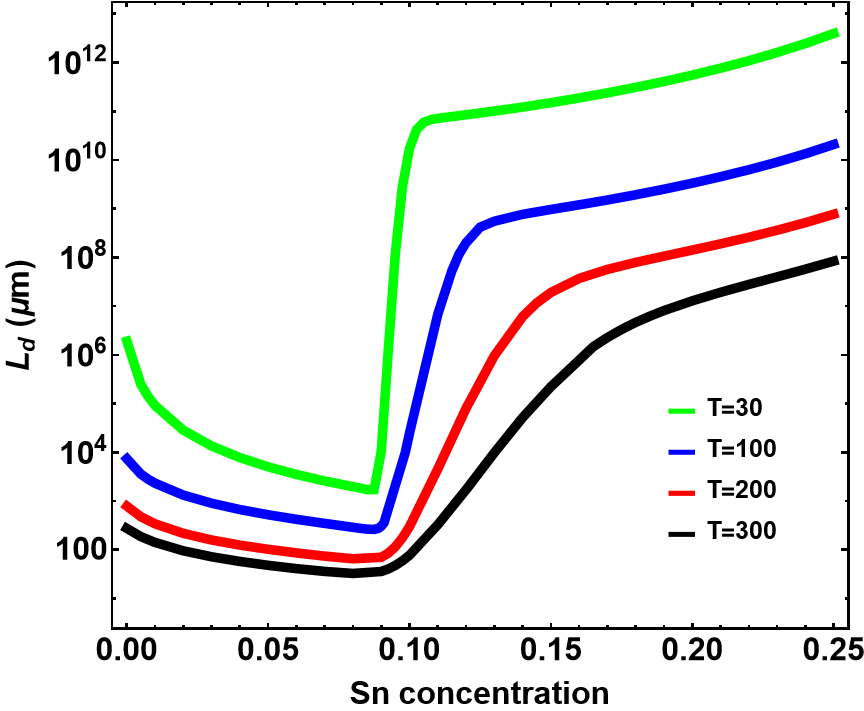}
    \caption{Spin diffusion length of GeSn (in $\mu$m) at temperatures T=300K (black), T=200K (red), T=100K (blue) and T=30K (green), as a function of Sn concentration.}
    \label{fig:difflength}
\end{figure}

Our spin-relaxation time was used, along with the previously calculated electron mobility \cite{sewell2025first} to calculate the spin diffusion length $L_d$, since $L_d=\mu_e \left< \tau \right> |\epsilon |$ for electric field strength $|\epsilon|$. The calculations were performed under an applied electric field of $1V$ across a material length of 1$\mu m$. Our results are shown in Figure \ref{fig:difflength}, where the spin diffusion length is given in micrometers as a function of Sn concentration at 300K, 200K, 100K, and 30K. In general, the behaviour of the spin diffusion length mirrors that of the spin-relaxation time and the mobility of GeSn, in that it decreases with Sn concentration until the indirect-to-direct bandgap transition, before increasing by several orders of magnitude. We calculate the room-temperature spin-diffusion length of Ge to be 275$\mu m$. At least $11\%$ Sn concentration is required to increase the spin diffusion length above that of Ge, while a Sn concentration of $25\%$ can increase the value to $100 m$. As temperature decreases, we achieve a higher spin-diffusion length, with Ge$_{0.75}$Sn$_{0.25}$ possessing a spin-diffusion length of the order of $10^{12} \mu m$ (1000km) at 30K. This large spin diffusion length can be attributed not just to the long spin-relaxation time, but also the high electron mobility of GeSn.

\section{Conclusion}

In conclusion, we use first-principles methods to calculate the spin-dependent alloy and e-ph scattering parameters of GeSn to determine the spin-relaxation time and diffusion length of GeSn as a function of alloy concentration, temperature and strain. We find that at low Sn concentrations, spin-alloy scattering reduces the spin-relaxation time. However, at higher Sn concentrations, the indirect-to-direct bandgap transition increases the spin-relaxation time by several orders of magnitude. Such increases are more pronounced at lower temperatures. At room temperature, the spin-relaxation time of GeSn is up to 1000 times that of Ge at Sn concentrations above $25\%$. The introduction of biaxial tensile strain reduces the Sn content required to achieve such an increase. The spin-relaxation time dramatically increases at the indirect-to-direct bandgap transition at cryogenic temperatures, requiring less Sn compared to room-temperature GeSn. For applications desiring a long spin-relaxation time, we propose an alloy of $15\%$ Sn above 50K, or $10\%$ Sn below 50K for unstrained GeSn. Our calculations also reveal a high spin-diffusion length in GeSn at compositions yielding a direct bandgap. Our model enables parameter transfer for multi-scale models. Compared to other direct-gap materials such as III-Vs, GeSn possesses a much longer spin-relaxation time. These findings highlight GeSn's potential as a promising material for spintronic applications.

\backmatter

\bmhead{Acknowledgements}

This work was supported by Science Foundation Ireland: Grant No. 19/FFP/6953.







\end{document}